\documentclass[aps,pra,reprint,a4paper]{revtex4-1}
\usepackage{amsmath}
\usepackage{amssymb}
\usepackage{graphicx}
\usepackage{color}
\usepackage{subfigure}
\usepackage[breaklinks]{hyperref}

\newcommand{\ket}[1]{\ensuremath{\left| #1 \right\rangle}}
\newcommand{\braket}[2]{\ensuremath{\left\langle #1 | #2 \right\rangle}}
\newcommand{\figref}[1]{Fig. \ref{#1}}

\begin{document}
\title{Signatures of the collapse and revival of a spin Schr\"{o}dinger cat state in a continuously monitored field mode}
\author{R.D. Wilson}
\author{M.J. Everitt}
\affiliation{Department of Physics, Loughborough University, Loughborough, Leicestershire LE11 3TU, UK}
\author{W.J. Munro}
\affiliation{NTT Basic Research Laboratories, NTT Corporation, 3-1 Morinosato-Wakamiya, Atsugi, Kanagawa 243-0198, Japan}
\author{Tim Spiller}
\affiliation{Quantum Information Science, School of Physics and Astronomy, University of Leeds, Leeds LS2 9JT, UK}
%\email{R.D.Wilson@lboro.ac.uk}
%\date{\today}

\begin{abstract}
	We study the effects of continuous measurement of the field mode during the collapse and revival of spin Schr\"{o}dinger cat states in the Tavis-Cummings model of $N$ qubits (two-level quantum systems) coupled to a field mode. We show that a compromise between relatively weak and relatively strong continuous measurement will not completely destroy the collapse and revival dynamics while still providing enough signal-to-noise resolution to identify the signatures of the process in the measurement record. This type of measurement would in principle allow the verification of the occurrence of the collapse and revival of a spin Schr\"{o}dinger cat state.
\end{abstract}

\maketitle

\section{Introduction}

	Two of the hallmarks of quantum behaviour, namely entanglement and macroscopically distinct superpositions of states (Schr\"{o}dinger cat states), are neatly encapsulated in the collapse and revival of Schr\"{o}dinger cat states in a coupled atom-field system \cite{Eberly1980,Gerry2004,Rodrigues2008,Jarvis2009,Jarvis2010,Everitt2012}. This interesting phenomenon also has a number of potential applications in quantum technologies, such as metrology or computing \cite{Munro2002,Ralph2003}, where the ability to swap a cat state between different subsystems would be an important resource.
	
	In the simplest case of a single qubit (two level quantum system) coupled to a field mode, the Jaynes-Cummings model \cite{Jaynes1963}, the qubit and field will initially entangle and $\left\langle\hat{\sigma}_{z}\right\rangle$ will begin to oscillate. The quantum information initially stored in the qubit is then transferred in to a Schr\"{o}dinger cat state of the field and the coherent oscillations collapse and the entanglement decays. Then the process essentially reverses, the field Schr\"{o}dinger cat state decays and the qubit and field reentangle, reviving the oscillations in $\left\langle\hat{\sigma}_{z}\right\rangle$. Given the field begins in a coherent state with an average of $\bar{n}$ photons, there are three timescales that characterise the collapse and revival dynamics: the Rabi time $t_{R}=\pi/\left(g\sqrt{\bar{n}}\right)$ which describes the underlying cyclical exchange of a photon; the collapse time $t_{c}=\sqrt{2}/g$ which sets the envelope for the Gaussian decay of the oscillations; and the first revival time $t_{r}=2\pi\sqrt{\bar{n}}/g$ when the oscillations reappear. When the system is extended to the case of $N$ qubits coupled to a field, the Tavis-Cummings model \cite{Tavis1968}, the characteristic timescales of the collapse and revival dynamics are altered. Importantly for our discussion the first revival time will now depend on the number of qubits and is given by $t_{r1}=t_{r}/N$ \cite{Jarvis2009}.	
	
	The effects of decoherence on the dynamics of collapse and revival of oscillations in a single qubit coupled to a field mode were studied in Ref.~\citenum{Everitt2009}. It was shown that as the level of decoherence in the field mode was increased there is a transition from the quantum limit of the field mode's behaviour, where we have collapse and revival of the oscillations in the qubit, to the classical limit of it's behaviour, where we simply see continuous Rabi oscillations in the qubit. When this approach was scaled up to study the collapse and revival of a spin Schr\"{o}dinger cat state in the Tavis-Cummings model it was found that increasing the number of spins in the system actually mitigates the effects of decoherence in the field and helps to preserve the collapse and revival behaviour \cite{Everitt2012}. One potential source of decoherence is quantum measurement, as the measurement interaction, whether it be discrete and projective or continuous, can be thought of as localising the quantum state in a similar way to other dissipative environments \cite{Joos1985,Wiseman2009}.
	
	Here we consider the situation where the decoherence in the field is the result of continuous measurement. We explore what trends can be seen in the record of this continuous measurement and discuss whether it may be possible to use these results to determine whether the collapse and revival process has taken place or not.

\section{Model}

	The Tavis-Cummings model consists of $N$ identical qubits which are coupled identically in the rotating wave approximation to a quantum field mode and can be described by a Hamiltonian of the form \cite{Tavis1968}
\begin{equation}
	H=\hbar \omega \hat{a}^{\dag}\hat{a}+\frac{\hbar\omega}{2}\sum_{k=1}^{N}\hat{\sigma}_{z}^{k}
	+\hbar g \sum_{k=1}^{N}\left(\hat{\sigma}_{+}^{k}\hat{a}+\hat{\sigma}_{-}^{k}\hat{a}^{\dag}\right).
	\label{eqn:TCHamiltonian}
\end{equation}
Here $\hat{a}^{\dag}\left(\hat{a}\right)$ are the creation (annihilation) operators of the field, $\sigma_{\pm}^{k}=1/2\left(\sigma_{x}^{k}\pm i\sigma_{y}^{k}\right)$ are the raising and lowering operators for the eigenstates of $\sigma_{z}^{k}$ of the $k^{\mathrm{th}}$ qubit and $g$ is the dipole coupling constant. Coherent states of the field mode will be given by
\begin{equation}
	\ket{\alpha}=\exp^{-\left|\alpha\right|^{2}/2}\sum_{n=0}^{\infty}\frac{\alpha^{n}}{n!}\left(\hat{a}^{\dag}\right)^{n}\ket{0}
	\label{eqn:coherent}
\end{equation}
where $\left|\alpha\right|^2$ is the mean number of photons. An analogue of these coherent states for a system of $N$ spins can also be defined 
\begin{equation}
	\ket{z,N}=\frac{1}{\left(1+\left|z\right|^{2}\right)^{N/2}}\bigotimes^{N}_{k=1}\left(\ket{e}_{k}+z\ket{g}_{k}\right)
	\label{eqn:spinCoherent}
\end{equation}
and these simply comprise a separable state of all spins pointing in the same direction \cite{Radcliffe1971,Arecchi1972,Zhang1990,Nemoto2000,Jarvis2009}. In further analogy with systems described in terms of position $\hat{q}$ and momentum $\hat{p}$ operators we can then define macroscopically distinct superpositions of these spin coherent states, or spin Schr\"{o}dinger cat states, of the form \cite{Gerry1997,Jarvis2010}
\begin{equation}
	\ket{\Theta_{\pm}\left(z,N\right)}=\frac{1}{\sqrt{2}}\left(\ket{z,N}\pm\ket{-z,N}\right).
	\label{eqn:spinCat}
\end{equation}
If a $N$ qubit system is prepared in such a Schr\"{o}dinger cat state and then allowed to interact with a field mode initialised in a coherent state. The collapse and revival dynamics will effectively swap the cat state into the field at $t=t_{r1}/2$ and then back into the spin subsystem at $t=t_{r1}$.

	The effects of continuous measurement on the evolution of the state vector \ket{\psi}, can be described by the quantum state diffusion formalism \cite{Percival1998};
\begin{multline}
	\ket{\mathrm{d}\psi}=-\frac{i}{\hbar}H\ket{\psi}\mathrm{d}t \\ +
	\left[\left\langle\hat{L}^{\dag}\right\rangle\hat{L}-\frac{1}{2}\hat{L}^{\dag}\hat{L} 
	-\frac{1}{2}\left\langle\hat{L}^{\dag}\right\rangle\left\langle\hat{L}\right\rangle\right]
	\ket{\psi}\mathrm{d}t \\ + 
	\left[\hat{L}-\left\langle\hat{L}\right\rangle\right]\ket{\psi}\mathrm{d}\xi ,
	\label{eqn:QSD}
\end{multline}
where $\ket{\mathrm{d}\psi}$ and $\mathrm{d}t$ are the state vector and time increments respectively, $\hat{L}$ is the Lindblad operator describing the measurement of the field and $\mathrm{d}\xi$ are the stochastic Wiener increments which satisfy $\overline{\mathrm{d}\xi^{2}}=\overline{\mathrm{d}\xi}=0$ and $\overline{\mathrm{d}\xi\mathrm{d}\xi^{*}}=\mathrm{d}t$. This stochastic evolution can be seen as a result of the random stream of measurement results recorded by an observer, known as the measurement record $r(t)$. Here we consider a measurement process described by the Lindblad $\hat{L}=\sqrt{2\Gamma}\hat{a}$ where an observer continuously monitors the position quadrature $\hat{q}=\sqrt{\hbar/\left(2m\omega\right)}\left(\hat{a}+\hat{a}^{\dag}\right)$ of the field mode with a measurement strength $\Gamma$ and extracts the measurement record \cite{Jacobs2006,Wiseman2009}
\begin{equation}
	\mathrm{d}r=\left\langle\hat{q}\right\rangle\mathrm{d}t+\frac{\mathrm{d}\xi}{\sqrt{8\Gamma}}.
	\label{eqn:xRecord}
\end{equation}

\section{Results}

	We consider a stochastic trajectory for a system with $N=5$ prepared initially in the state $\ket{\alpha=\sqrt{25}}\otimes\ket{\Theta\left(z=1,N=5\right)}$. Time-frequency analysis is performed using a continuous wavelet transform with a Morlet wavelet to study the trends in the measurement record as a function of the measurement strength. 

%\begin{widetext}
\begin{figure}
	\begin{center}
	\includegraphics[width=0.49\textwidth,trim=2.5mm 6mm 6mm 4mm,clip]{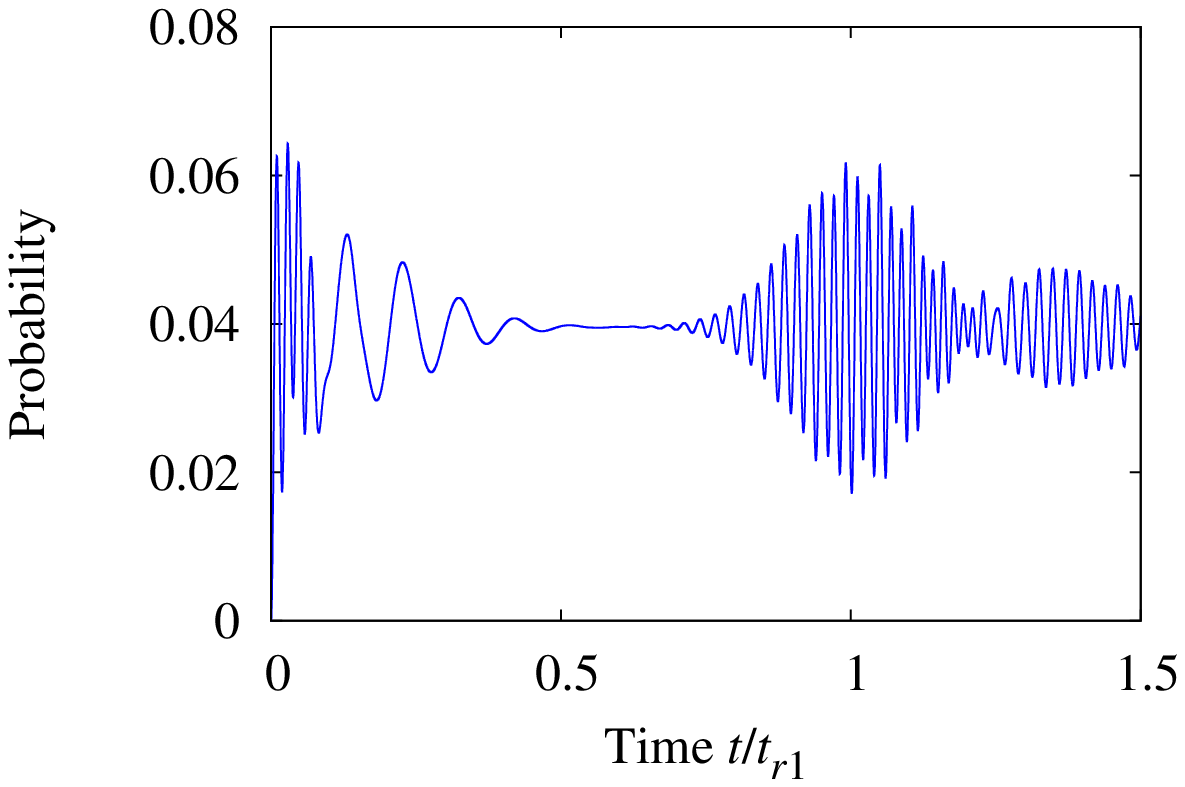}
	\hbox{\hspace{6ex}\includegraphics[width=0.445\textwidth,trim=5mm 2mm 15mm 9.5mm,clip]{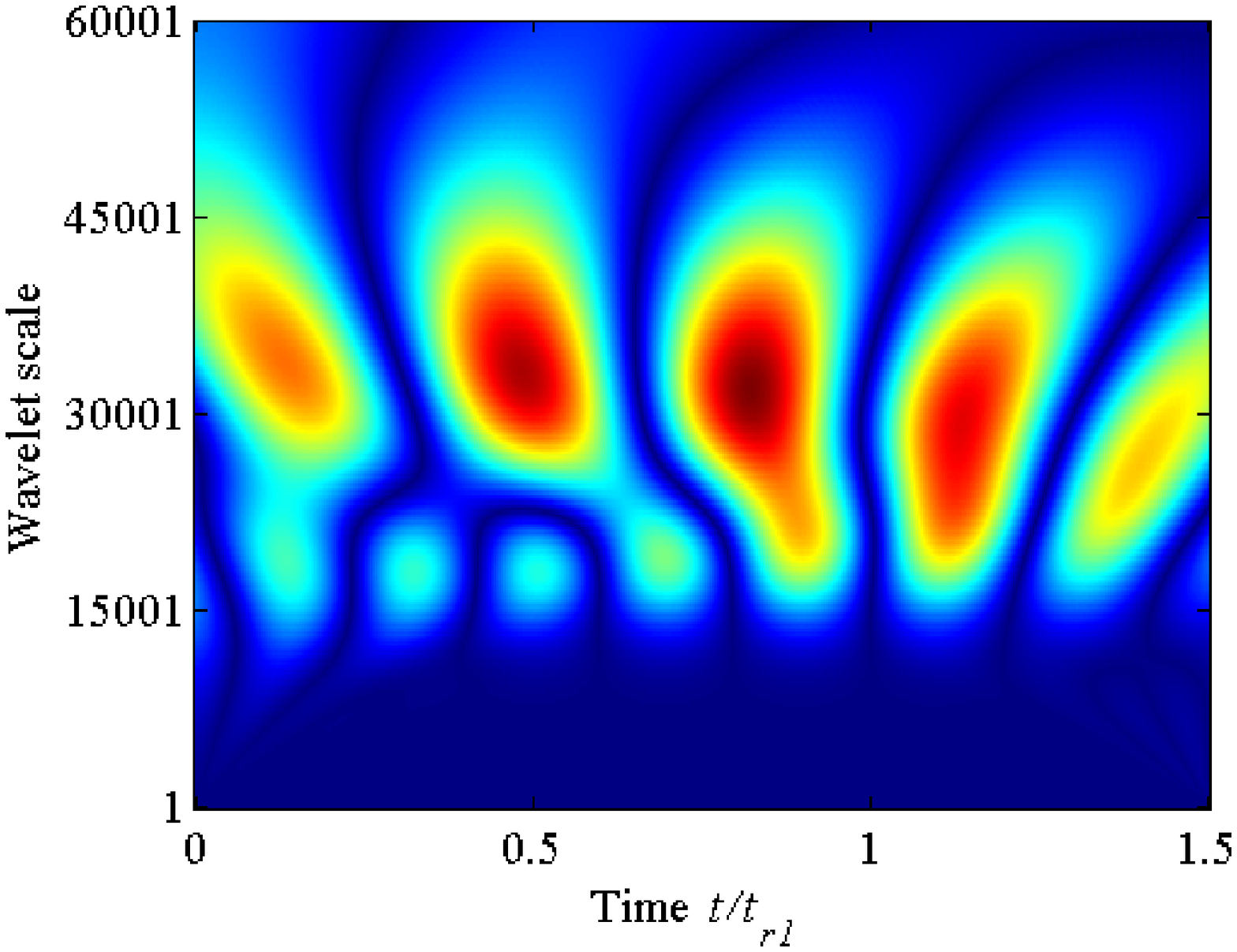}}
	\caption{(Colour online) Collapse and revival dynamics of the Tavis-Cummings system in the absence of measurement ($\gamma=0$) for an initial state of $\ket{\alpha=\sqrt{25}}\otimes\ket{\Theta\left(z=1,N=5\right)}$. The top pane shows the characteristic collapse and revival of oscillations in the probability of finding all the qubits in their ground state $\left|\braket{\psi}{ggggg}\right|^{2}$. The bottom pane shows the continuous wavelet transform of the time integrated position expectation value for the field mode $r_{I}(t)$, that is the measurement record in the idealised case of zero back-action and infinite signal-to-noise ratio. The wavelet transform is normalised across all wavelet scales $a$ with red denoting coefficients with the highest power and blue the lowest.}
	\label{fig:SEPlots}
	\end{center}
\end{figure}
%\end{widetext}

%\begin{widetext}

\begin{figure*}[p!]
	\begin{center}
	\subfigure[\ $\Gamma=0.5\times 10^{-5}$]{\includegraphics[width=0.355\textwidth,trim=2.5mm 6mm 6mm 4mm,clip]{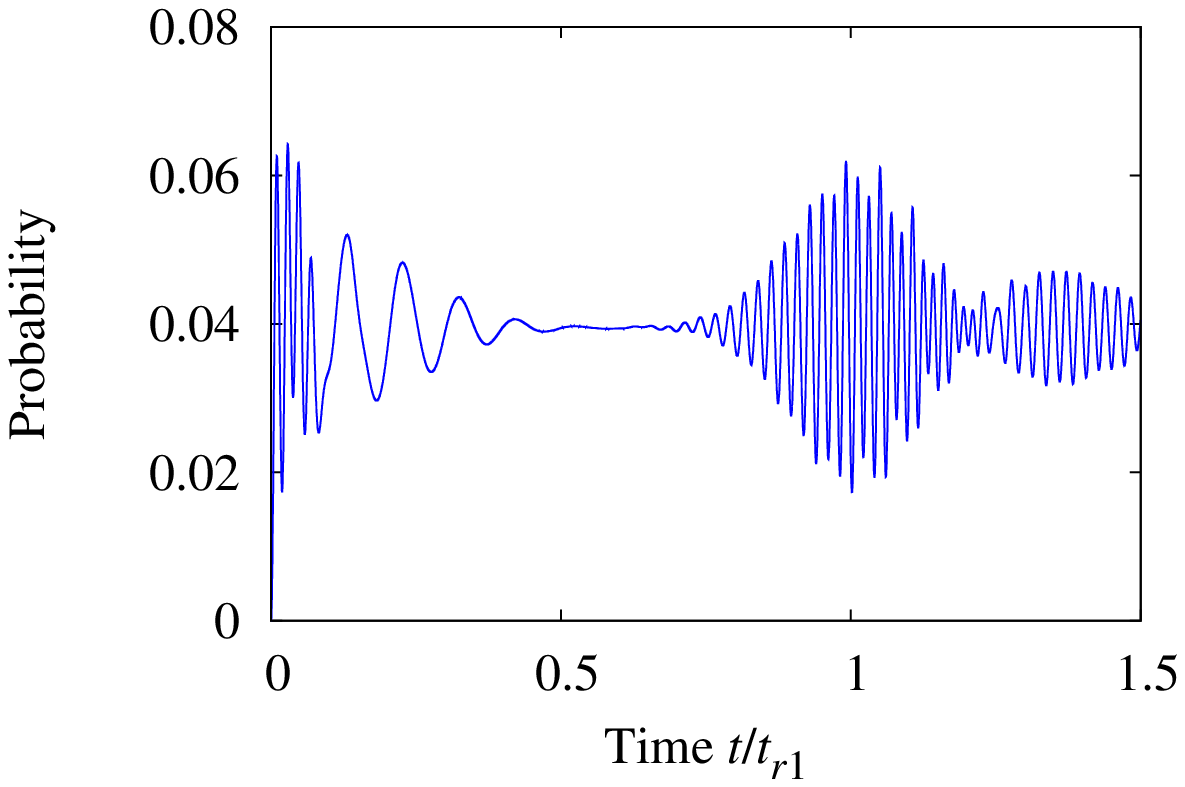}
	\includegraphics[width=0.3075\textwidth,trim=5mm 2mm 15mm 9.5mm,clip]{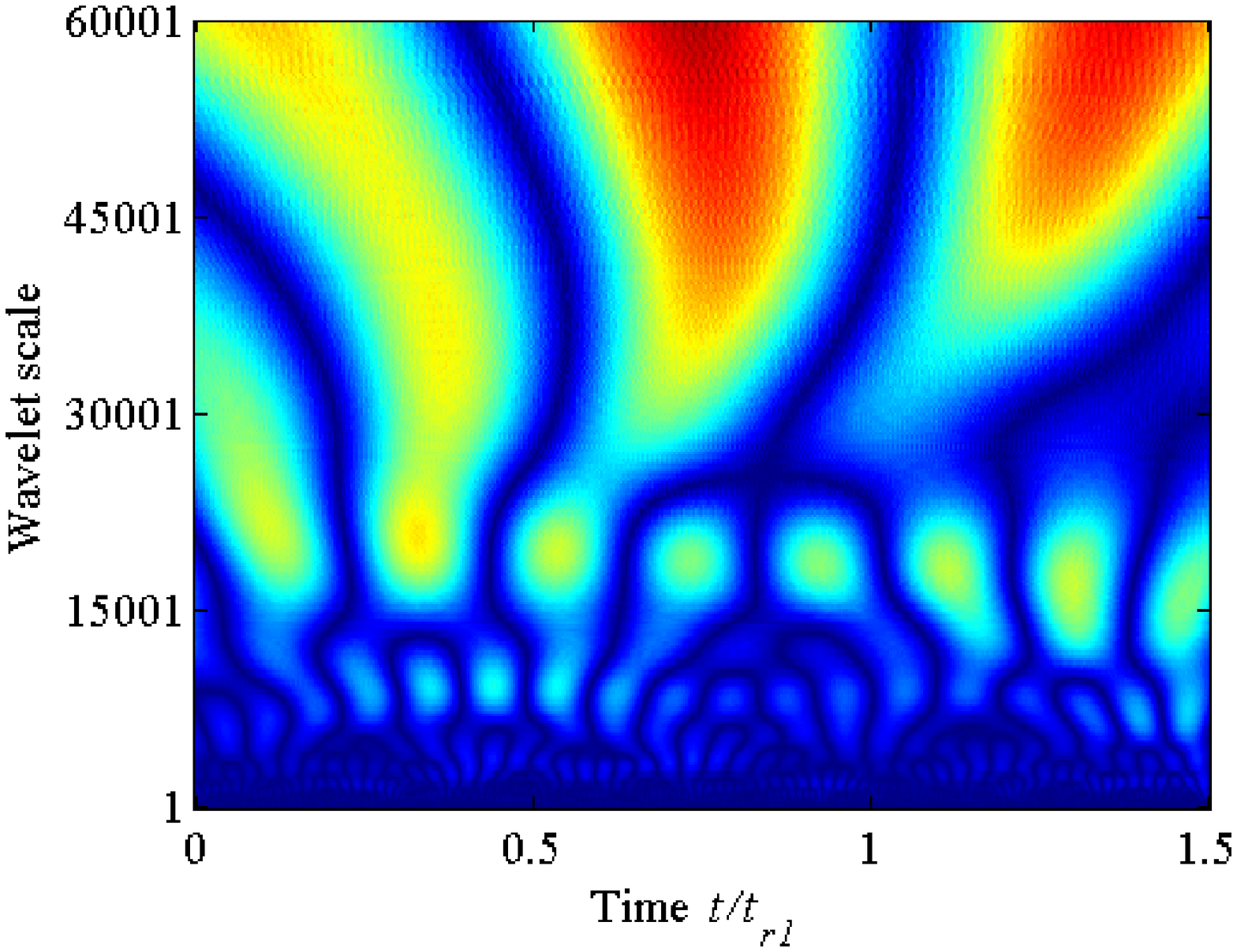}}
	
	\subfigure[\ $\Gamma=0.5\times 10^{-4}$]{\includegraphics[width=0.355\textwidth,trim=2.5mm 6mm 6mm 4mm,clip]{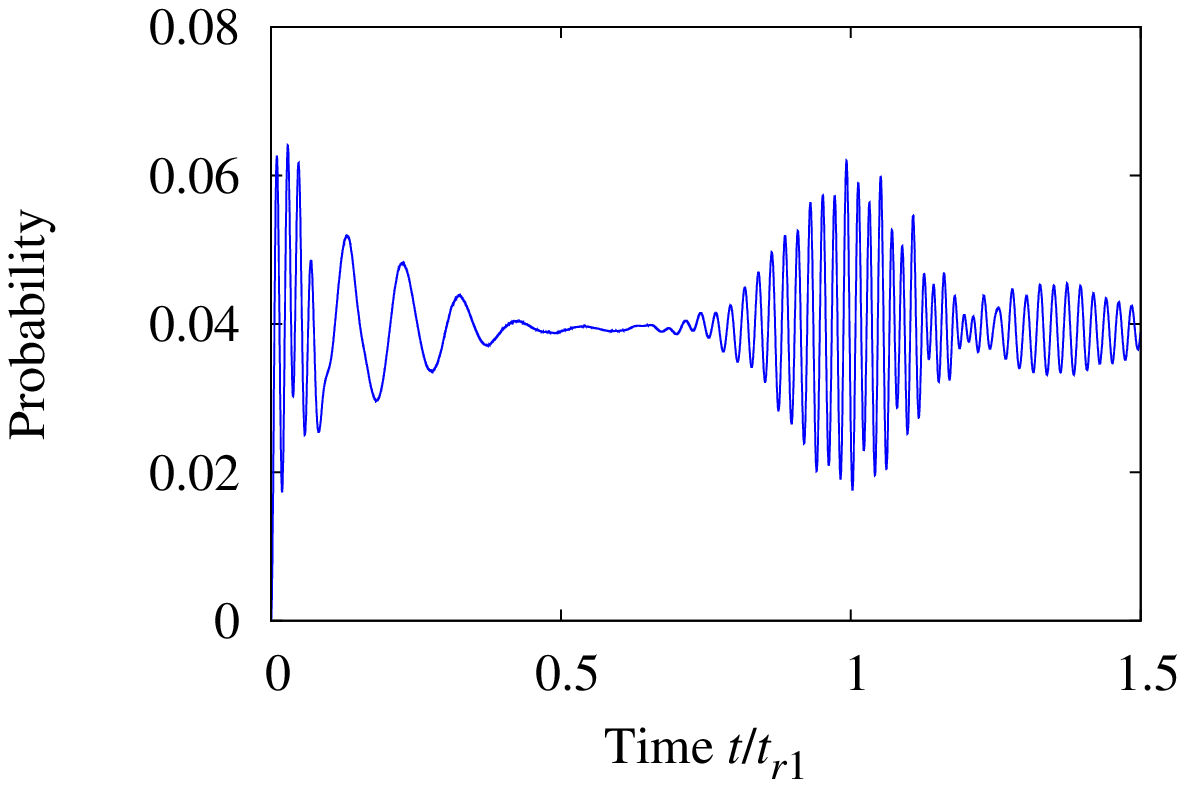}
	\includegraphics[width=0.3075\textwidth,trim=5mm 2mm 15mm 9.5mm,clip]{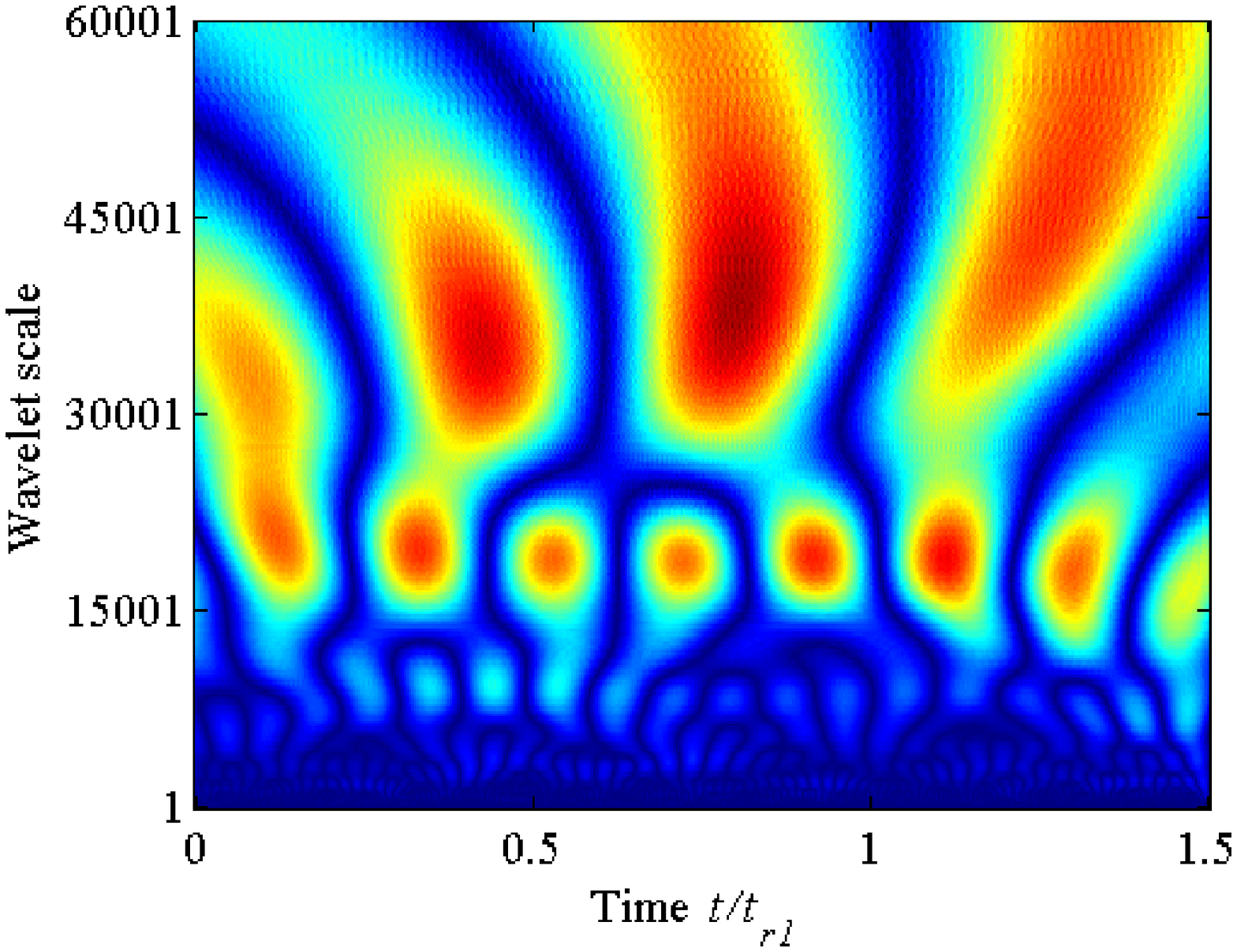}}
	
	\subfigure[\ $\Gamma=0.5\times 10^{-3}$]{\includegraphics[width=0.355\textwidth,trim=2.5mm 6mm 6mm 4mm,clip]{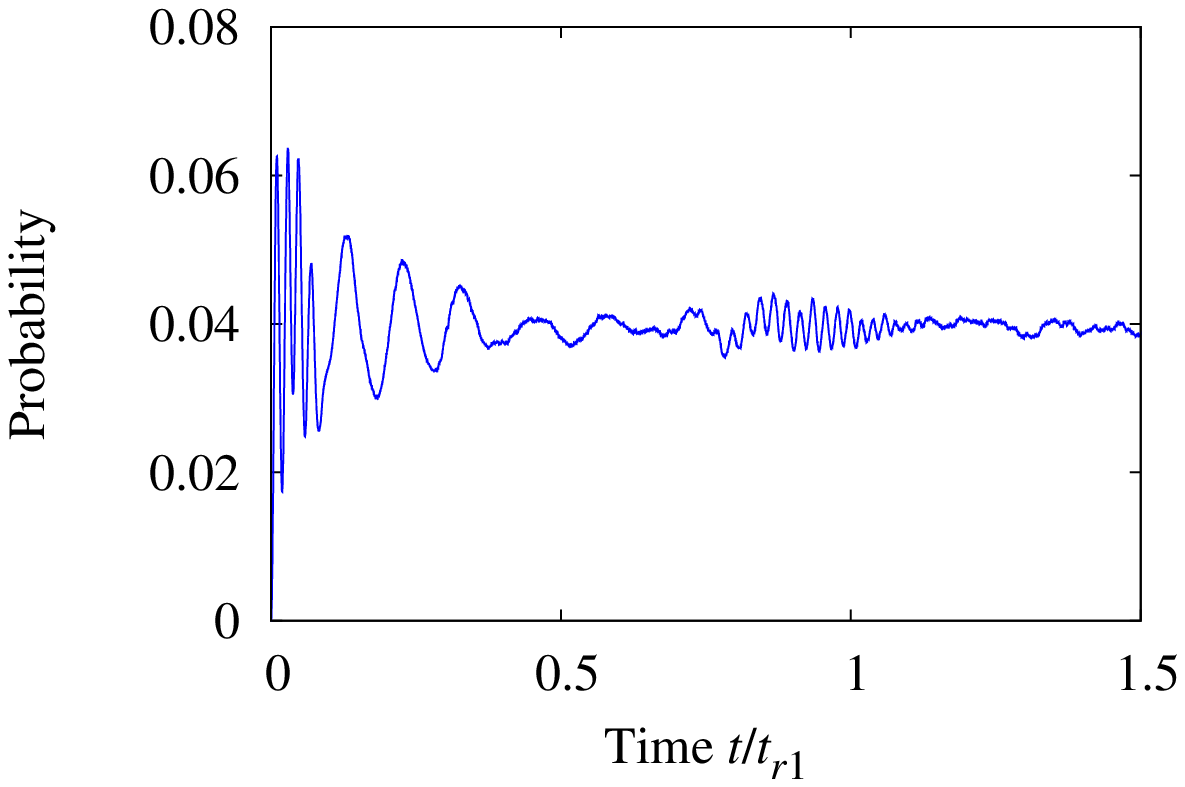}
	\includegraphics[width=0.3075\textwidth,trim=5mm 2mm 15mm 9.5mm,clip]{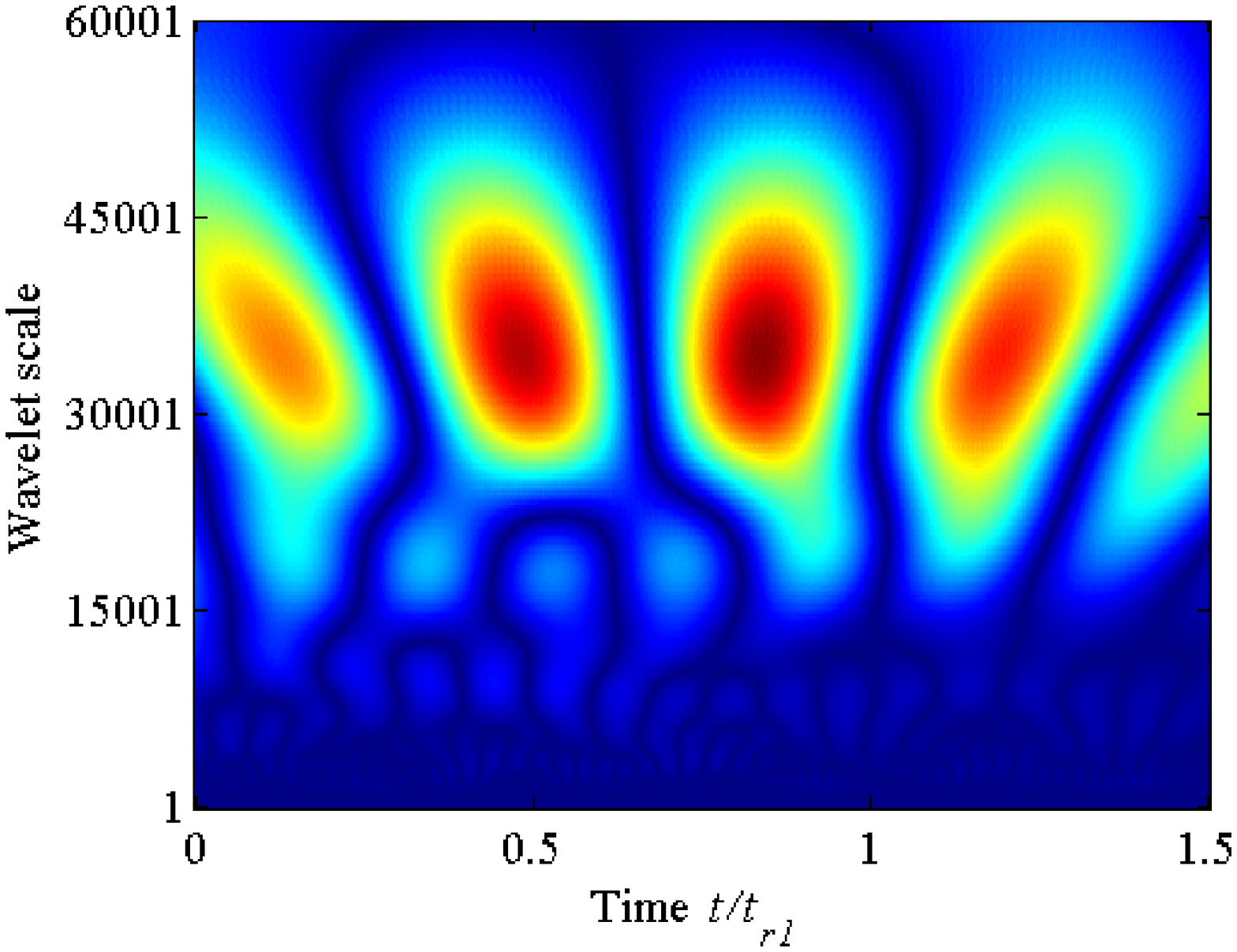}}
	
	\subfigure[\ $\Gamma=0.5\times 10^{-2}$]{\includegraphics[width=0.355\textwidth,trim=2.5mm 6mm 6mm 4mm,clip]{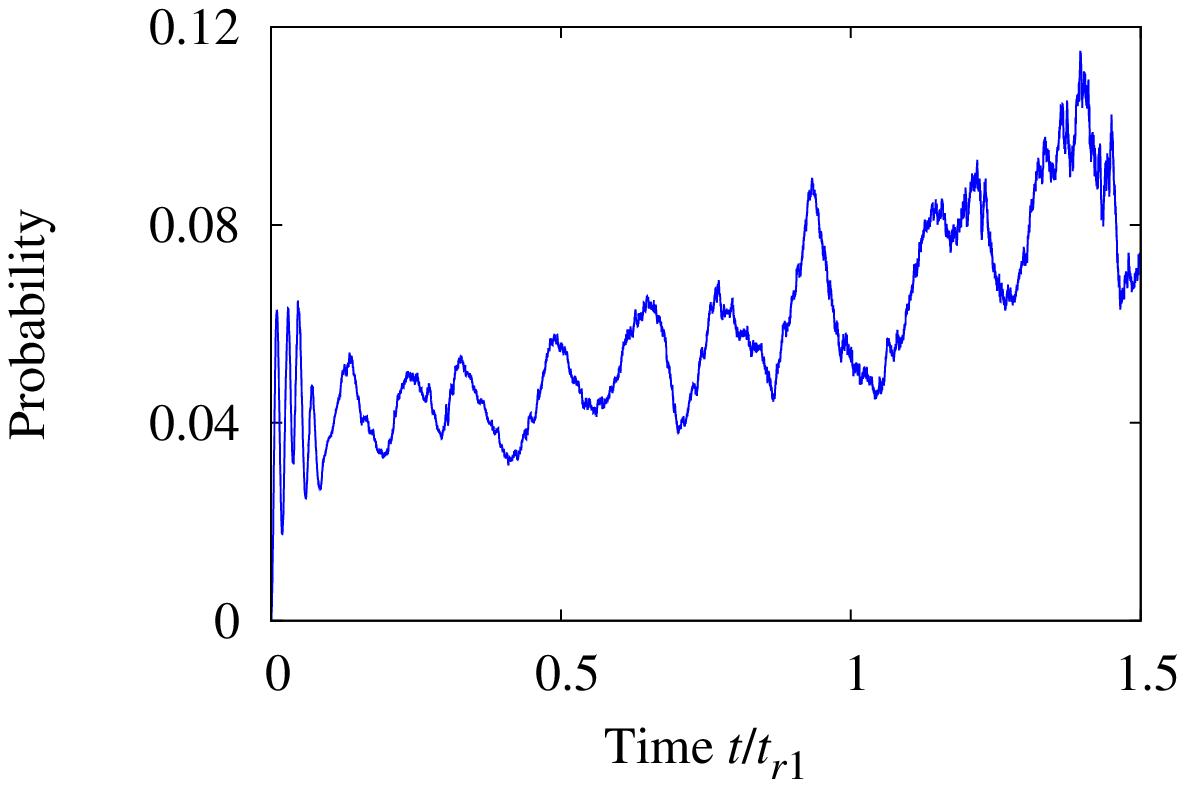}
	\includegraphics[width=0.3075\textwidth,trim=5mm 2mm 15mm 9.5mm,clip]{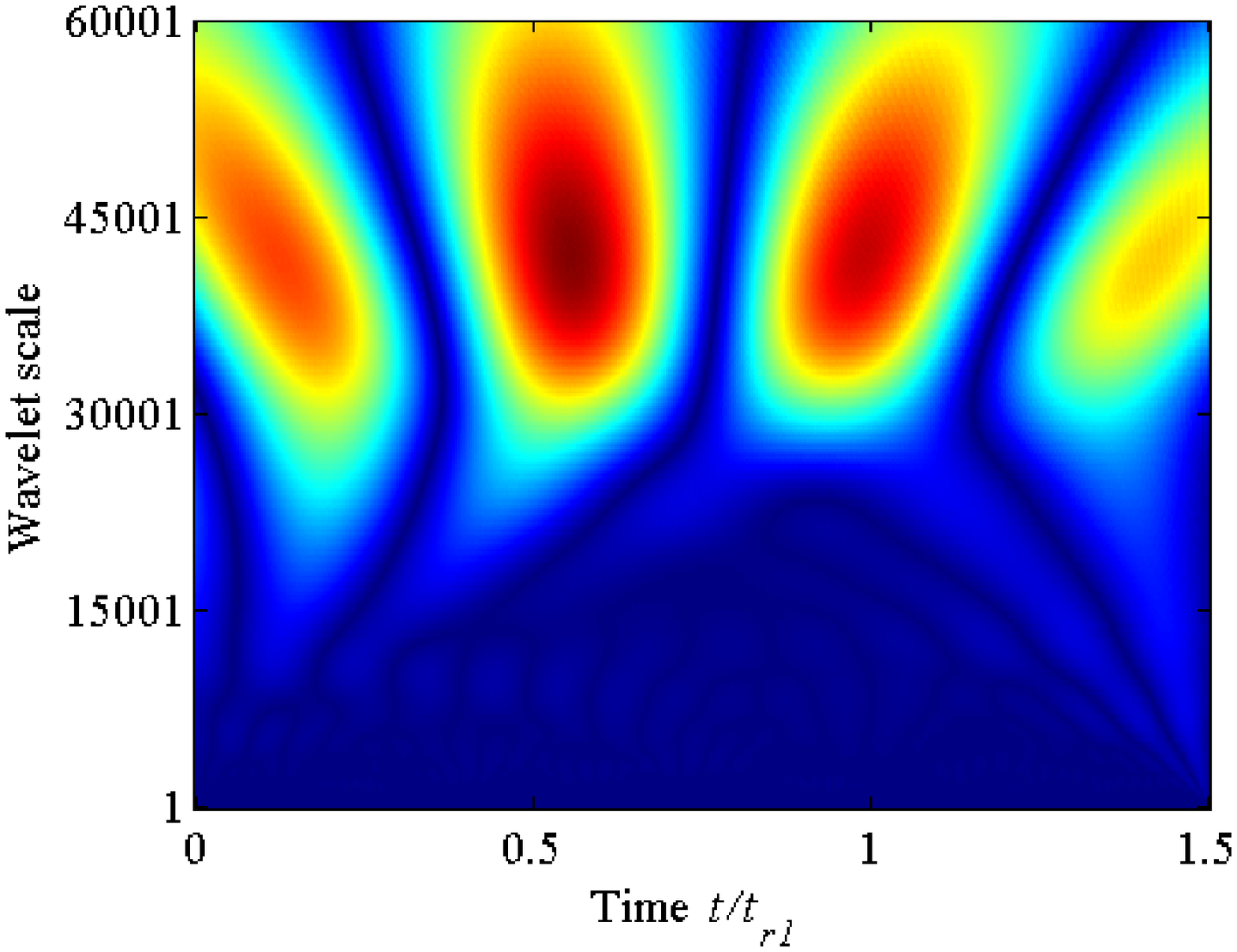}}
	\caption{(Colour online) Collapse and revival dynamics of the Tavis-Cummings system as a function of the measurement strength $\Gamma$ for a single stochastic trajectory with an initial state of $\ket{\alpha=\sqrt{25}}\otimes\ket{\Theta\left(z=1,N=5\right)}$. For each row: the left pane shows the time evolution of the probability of finding all the qubits in their ground state $\left|\braket{\psi}{ggggg}\right|^{2}$; and the left pane shows the continuous wavelet transform of the measurement record extracted from the field mode $r(t)$. The wavelet transform is normalised across all wavelet scales $a$ with red denoting coefficients with the highest power and blue the lowest.}
	\label{fig:QSDPlots}
	\end{center}
\end{figure*}

%\end{widetext}
%\vspace{10mm}

	We begin by studying the dynamics in the idealised case of the absence of measurement induced decoherence, where $\Gamma=0$ and \eqref{eqn:QSD} essentially reduces to the Schr\"{o}dinger equation. In \figref{fig:SEPlots} we can clearly see the characteristic collapse and revival of oscillations in the qubit state. The the time integrated position operator expectation value, given by the first term in \eqref{eqn:xRecord}, can be interpreted as the ideal measurement record $r_{I}(t)$. That is, the record of a measurement with zero back action on the system and with an infinite signal-to-noise ratio. The continuous wavelet transform for this ideal measurement record is also shown in \figref{fig:SEPlots} and can be used as a benchmark for comparison of  the key trends in the collapse and revival measurement record. We note that the collapse and revival process over the first revival time is characterised by a series of high scale (low frequency) nodes.
	 
	The dynamics of a single stochastic trajectory for a range of increasing measurement strengths $\Gamma$ are shown in \figref{fig:QSDPlots}. For relatively weak measurements, \figref{fig:QSDPlots}~(a), we can see that the very low signal-to-noise ratio of the measurement record means that the characteristic low frequency signatures of the collapse and revival process are completely obscured by noise. This is despite the fact that the measurement process at this value of $\Gamma$ causes relatively little disturbance to the collapse and revival dynamics. At the other end of the scale, for relatively strong measurements, \figref{fig:QSDPlots}~(d), we can see that although the measurement record has a comparatively good signal-to-noise ratio the underlying dynamics of the system have been disturbed so much that the collapse and revival process and it's corresponding time-frequency signatures are essentially destroyed.

	The middle ground of measurement strengths in the range $0.5\times 10^{-5}<\Gamma<0.5\times 10^{-2}$ appears to offer a better compromise. Here the measurement is weak enough to not disturb the collapse and revival process too much, while still providing a reasonable signal-to-noise ratio. In \figref{fig:QSDPlots}~(c), the characteristic structure of nodes in the low frequency range of the wavelet transform caused by the collapse and revival dynamics, as seen in \figref{fig:SEPlots}, are reproduced to a good approximation, clearly indicating that collapse and revival has taken place. This occurs despite the fact that the measurement back action has somewhat damped the revival of oscillations in the qubit dynamics. Where as in \figref{fig:QSDPlots}~(b), the qubit dynamics are a better approximation of the collapse and revival dynamics in the ideal case. The lower signal-to-noise ratio means that the low frequency structure in the measurement record is somewhat obscured, although there is still some correlation.

\section{Conclusions}

	To conclude, we have investigated the effects of continuous measurement of the field during the collapse and revival of a spin Schr\"{o}dinger cat state in the Tavis-Cummings model and discussed whether the characteristic signatures of the collapse and revival dynamics can be identified in the record of this measurement. We find that relatively strong measurements destroy the collapse and revival dynamics, where as relatively weak measurements don't allow enough signal-to-noise resolution to identify the characteristic low frequency structure in the measurement record. The middle ground between these two regimes offers the most promising results as the disturbance to the collapse and revival process caused by the measurement back-action is still relatively small, while the improved signal-to-noise resolution allows the low frequency signatures in the measurement record to be identified. This type of continuous measurement of the field should allow the occurrence of the collapse and revival process in the system of interest to be verified.

\section*{Acknowledgements}

	R.D.W. and M.J.E. acknowledge that this publication was made possible through the support of a grant from the John Templeton Foundation.

%\bibliography{crmRefs}

%merlin.mbs apsrev4-1.bst 2010-07-25 4.21a (PWD, AO, DPC) hacked
%Control: key (0)
%Control: author (8) initials jnrlst
%Control: editor formatted (1) identically to author
%Control: production of article title (-1) disabled
%Control: page (0) single
%Control: year (1) truncated
%Control: production of eprint (0) enabled
%

\end{document}